\documentclass[aps,nofootinbib]{revtex4-1}
\pdfoutput=1
\usepackage{graphicx,subfigure}
\usepackage{amsmath,amssymb}
\usepackage{color}
\usepackage{mathrsfs}

\usepackage{graphicx,subfigure}

\def\beq{\begin{equation}}
\def\eeq{\end{equation}}
\def\beqr{\begin{eqnarray}}
\def\eeqr{\end{eqnarray}}
\def\bdpm{\begin{displaymath}}
\def\edpm{\end{displaymath}}
\def\half{\frac{1}{2}}
\definecolor{lgray}{gray}{0.6}

\newcommand{\tbt}{t_\beta}
\newcommand{\tth}{t_\theta}
\newcommand{\cth}{c_\theta}
\newcommand{\sth}{s_\theta}

\newcommand{\msbar}{\overline{\textrm{MS}}}
\newcommand{\Veff}{V_\textrm{eff}}

\begin{document}

\title{
Scalar dark matter in the conformally invariant extension 
of the standard model
}

\author{Dong-Won Jung}
\email{dongwonj@korea.ac.kr}

\author{Jungil Lee}
\email{jungil@korea.ac.kr}

\author{Soo-hyeon Nam}
\email{glvnsh@gmail.com}

\affiliation{ 
Department of Physics, 
Korea University, Seoul 02841, Republic of Korea
}

\date{\today}

\begin{abstract}

We study a classically scale-invariant model 
with an electroweak singlet scalar mediator together with a scalar dark matter multiplet of global $O(N)$ symmetry. 
Our most general conformally invariant scalar potential
generates the electroweak symmetry breaking via the Coleman-Weinberg mechanism,
and the new scalar singlet acquires its mass through radiative corrections
of the SM particles and the scalar dark matter.
Taking into account the collider bounds, 
we present the allowed region of new physics parameters satisfying the recent measurement of relic abundance.
With the obtained parameter sets, 
we predict the elastic scattering cross section of the new scalar multiplet
into target nuclei for a direct detection of the dark matter. 
We also perform a full analysis with arbitrary set of parameters for $N \geq 2$,
and discuss the implication of the constraints from the on-going direct and indirect detections of dark matter.

\end{abstract}

\maketitle

\section{Introduction}

Since the discovery of the Higgs boson at the Large Hadron Collider (LHC) 
\cite{Aad:2012tfa,Chatrchyan:2012xdj}, the Standard Model (SM) has been
remaining quite successful in spite of  many experimental pursuits of physics
beyond the SM (BSM). Nevertheless, the story has not been completed yet: Higgs self coupling $\lambda_h$
should be measured to prove or disprove whether the Higgs mechanism is the root cause of
 the Electroweak Symmetry Breaking (EWSB) or not. Besides this lack of the experimental 
confirmation of the EWSB mechanism, there are still many open questions  
both theoretically and phenomenologically. Among them are the gauge hierarchy problem that 
addresses the smallness of the EWSB scale compared to the Planck scale $M_P$, 
and non-baryonic dark matter (DM) of the Universe.

In 1995, Bardeen \cite{Bardeen:1995kv} suggested that softly broken conformal (scale) 
invariance could be a possible solution for the hierarchy problem. 
If conformal invariance is assumed at classical level, there is no dimensionful parameter in the Lagrangian 
and conformal symmetry is broken only logarithmically through the conformal anomaly
with dimension-4 operators at quantum level. 
Many works have been done along this line with the EWSB by dimensional transmutation    
\cite{Hur:2007uz,
Ko:2008ug,
Ko:2009zz,
Ko:2010rj,
Hur:2011sv,
Heikinheimo:2013fta,
Heikinheimo:2013xua,
Holthausen:2013ota,
Jung:2014zga,
Salvio:2014soa,
Kubo:2014ova,
Kubo:2014ida,
Schwaller:2015tja,
Ametani:2015jla,
Kubo:2015joa,
Haba:2015qbz,
Hatanaka:2016rek,
Kubo:2016kpb,
Kubo:2018vdw}, 
or radiative mechanisms 
\cite{Meissner:2006zh,
Foot:2007as,
Meissner:2007xv,
Hambye:2007vf,
Foot:2007iy,
Iso:2009ss,
Iso:2009nw,
Holthausen:2009uc,
AlexanderNunneley:2010nw,
Ishiwata:2011aa,
Lee:2012jn,
Okada:2012sg,
Iso:2012jn,
Gherghetta:2012gb,
Das:2011wm,
Carone:2013wla,
Khoze:2013oga,
Farzinnia:2013pga,
Gabrielli:2013hma,
Antipin:2013exa,
Hashimoto:2014ela,
Hill:2014mqa,
Guo:2014bha,
Radovcic:2014rea,
Binjonaid:2014oga,
Davoudiasl:2014pya,
Allison:2014zya,
Farzinnia:2014xia,
Pelaggi:2014wba,
Farzinnia:2014yqa,
Foot:2014ifa,
Benic:2014aga,
Guo:2015lxa,
Oda:2015gna,
Das:2015nwk,
Fuyuto:2015vna,
Endo:2015ifa,
Endo:2015nba,
Plascencia:2015xwa,
Hashino:2015nxa,
Karam:2015jta,
Ahriche:2015loa,
Wang:2015cda,
Haba:2015nwl,
Ghorbani:2015xvz,
Helmboldt:2016mpi,
Jinno:2016knw,
Ahriche:2016cio,
Ahriche:2016ixu,
Das:2016zue,
Khoze:2016zfi,
Karam:2016rsz,
Oda:2017kwl,
Ghorbani:2017lyk,
Brdar:2018vjq,
Brdar:2018num,
YaserAyazi:2018lrv,
YaserAyazi:2019caf,
Mohamadnejad:2019wqb,
Kim:2019ogz} of Coleman-Weinberg (CW) \cite{Coleman:1973jx} types. 
In many of those works, the DM models are embedded at the same time. Incorporating the DM 
in a conformally invariant setup is highly motivated since in both cases BSM 
particles are required. For example, a naive Coleman-Weinberg type of the model suffers instability 
arising from radiative corrections because of the large top-quark mass, so it should be 
stabilized with the introduction of additional bosonic degrees of freedom. 
As one of the simplest extensions of the SM, 
one can introduce a gauge-singlet Higgs-portal scalar DM, 
and make it stabilize the Higgs mass via the CW mechanism 
as shown in Refs.~\cite{Endo:2015ifa,Endo:2015nba}.
However, generating a proper Higgs mass requires a large enough mass of the scalar DM, 
so that the Higgs-scalar couplings become too large. 
As a result, this kind of a simple setup makes the theory non-perturbative at a few TeV scale, which is undesirable.  
This issue can be resolved by separating the new scalar responsible for the EWSB 
from the DM sector.

In this work, we propose the conformally invariant DM model which 
contains an $SU(2)$ doublet Higgs field, a scalar mediator together with 
hidden sector scalar particles with $O(N)$ global symmetry for the stability of 
the hidden sector. With the most general conformally invariant Lagrangian, 
we employ the framework of Gildener-Weinberg (GW) \cite{Gildener:1976ih}, 
where a flat direction at tree-level is lifted by radiative corrections. 
In this way, the vacuum structure is determined and a light pseudo Nambu-Goldstone boson,
called `scalon', appears as a result of conformal symmetry breaking. 
Through its mixing with the SM-like Higgs boson, two light scalar particles emerge in the model 
that interact with both visible and hidden sectors. 
In particular, there exist contact interactions of the DMs and, as a result, 
the model takes the form of so-called the `secluded' DM \cite{Pospelov:2007mp} scenarios
as a conformally invariant version.

This paper is organized as follows. In Sec.\ II,  we describe the model 
 in detail including the determination of the flat direction in a similar manner to the GW framework. 
Next, the effective potential obtained by the CW mechanism 
and the radiatively induced scalar masses are presented at one-loop level in Sec.\ III. 
In Sec.\ IV, we provide a detailed phenomenological analysis of the DM physics
 such as the relic density and the direct detection. 
Section\ V is devoted to a summary and the conclusion.

\section{model}

We consider a dark sector consisting of two classically massless real scalar fields $S$ and $\phi$, 
which are SM gauge singlets.  
The scalar mediator $S$ is responsible for the EWSB together with the SM Higgs doublet $H$, 
and the scalar $\phi$ is a DM candidate chosen to be the fundamental representation of a global $O(N)$ group, 
$\phi = (\phi_1, \cdots, \phi_N)^T$. 
The extended Higgs sector Lagrangian with the renormalizable DM interactions is then given by
\beq \label{eq:DM_Lagrangian}
\mathscr{L}_{\rm DM} = \left(D_\mu H\right)^{\dagger}D^\mu H 
	+ \half\left(\partial^{\mu} S\right)^2 +\half(\partial_\mu\phi^T)\partial^\mu\phi - V(H, S, \phi), 
\eeq
where the scale-invariant scalar potential is
\beq \label{eq:VS_potential}
 V(H, S, \phi) =  \lambda_h (H^{\dagger} H)^2 + \half\lambda_{hs} H^{\dagger} H S^2  
 	+ \half\lambda_{h\phi} H^{\dagger} H \phi^T\phi +\frac{1}{4}\lambda_{s\phi} S^2\phi^T\phi 
 	+ \frac{1}{4}\lambda_s S^4  + \frac{1}{4}\lambda_\phi (\phi^T\phi)^2.
\eeq
A similar model was studied recently for $N=2$ case in Refs. \cite{Ghorbani:2015xvz, Ghorbani:2017lyk},
but they simply assumed $\lambda_{h\phi} = 0$ in order to decouple the DM sector from the SM Higgs.
In general, however, such an interaction term is not forbidden by a discrete symmetry such as $Z_2$ symmetry theoretically,
and also it is very important to explain the current astronomical observables phenomenologically,
as we will show in Sec.\ V.

After the EWSB, the DM scalar $\phi$ takes a vanishing VEV  
while the neutral component of the SM Higgs and the singlet scalar $S$ develop nonzero VEVs, 
$\langle H^0 \rangle=v_h/\sqrt{2}$ and $\langle S \rangle = v_s$, respectively.
Adopting the GW approach, we choose a flat direction among the scalar VEVs
along which the potential in Eq.~(\ref{eq:VS_potential}) vanishes at some scale $\mu = \Lambda$. 
Along that flat direction, the potential minimization conditions
$\partial V/\partial H |_{\langle H^0\rangle=v_h/\sqrt{2}} = \partial V/\partial S |_{\langle S\rangle=v_s}=0$
lead to the following relations
\beq \label{eq:lambdamin}
\lambda_{hs}(\Lambda) = - 2\lambda_h(\Lambda) /\tbt^2 , \qquad 
\lambda_s(\Lambda) = \lambda_h(\Lambda) / \tbt^4 ,
\eeq
where $\tbt\, (\equiv \tan\beta) = v_s/v_h$.
The neutral scalar fields $h$ and $s$ defined by $H^0=(v_h+h)/\sqrt{2}$ and $S=v_s+s$ 
are mixed to yield the mass matrix:
\beq \label{eq:mass_mattrix}
\mu_{h}^2 = 2 \lambda_h v_h^2 , \qquad
\mu_{s}^2 = 2 \lambda_h v_h^2 /\tbt^2 , \qquad
\mu_{hs}^2 = - 2 \lambda_h v_h^2 /\tbt.
\eeq
The corresponding scalar mass eigenstates $h_1$ and $h_2$ are admixtures of $h$ and $s$:
\beq
\left( \begin{array}{c} h_1 \\[1pt] h_2 \end{array} \right) =
\left( \begin{array}{cc} \cos \theta &\ -\sin \theta \\[1pt]
	 \sin \theta &\ \cos \theta \end{array} \right)
\left( \begin{array}{c} h \\[1pt] s \end{array} \right) ,
\eeq
where the mixing angle $\theta$ is given by
\beq \label{eq:tan_theta}
\tan \theta = \frac{y}{1+\sqrt{1+y^2}}, \qquad y \equiv  \frac{-2 \mu_{hs}^2}{\mu_h^2 - \mu_s^2}.
\eeq
Combining Eqs.~(\ref{eq:mass_mattrix}) and (\ref{eq:tan_theta}), 
we have $\tan\theta = -t_\beta$ or $1/t_\beta$.
The mixing angle $\theta$ is expected to be very small (less than about 0.2) due to the LEP constraints.
If $\tan\theta = -t_\beta$, then $\lambda_s = \lambda_h/\tan^4\!\theta$ from Eq.~(\ref{eq:lambdamin})
so that $\lambda_s$ becomes very large.  
But this case is theoretically disfavored because of the failure of the perturbativity of the couplings.
Also, experimental constraints disfavor this scenario as well \cite{Farzinnia:2013pga}. 
Therefore, we only consider the case of $\tan\theta\, (\equiv \tth) = 1/t_\beta$,
which results in $\sin\theta\, (\equiv s_\theta) = c_\beta$ and $\cos\theta\, (\equiv c_\theta) = s_\beta$.
In this case, $\lambda_{hs}$ and $\lambda_s$ are suppressed by $\tth^2$ and $\tth^4$, respectively, 
which ensures the perturbativity of those couplings 
and induces the suppression of the Higgs portal interactions.  
As a result, the scalar couplings in Eq.~(\ref{eq:VS_potential}) change very slowly with $\Lambda$,
and similar discussions can be found also in Refs. \cite{Gildener:1976ih, Hempfling:1996ht}.  

After diagonalizing the mass matrix, 
we obtain the physical masses of the two scalar bosons ($h_1, h_2$) and the DM scalar $\phi$ as follows:
\beq \label{eq:scalar_mass_tree}
M^2_1 = 2 \lambda_h v^2 \tth^2, \quad M^2_2 = 0,
	\quad M_{\phi}^2 = \frac{v^2}{2}\left(\lambda_{h\phi}\sth^2 + \lambda_{s\phi}\cth^2\right),
\eeq
where $v\ (\equiv \sqrt{v_h^2 + v_s^2})$ can be considered to be the VEV of the radial component of a scalar field composed of $h$ and $s$.
The value of $v$ is determined from the radiative corrections 
and is set to be the scale about $\Lambda$ according to GW. 
We assume that $M_1$ corresponds to the observed SM-like Higgs boson mass in what follows.
The SM Higgs $h_1$ and the DM scalars $\phi$ have the tree-level masses 
while the new scalar singlet $h_2$ acquires its mass through radiative corrections, 
which is similar to the cases considered in Refs. \cite{Farzinnia:2014yqa, Ghorbani:2015xvz}. 
In terms of the physical states, the tree-level scalar potential in the flat direction can be expressed 
in the unitary gauge as
\beqr \label{eq:V_interaction}
 V(h_1, h_2, \phi_i) &=& \half M_1^2 h_1^2 + \half M_\phi^2 \phi_i^2 +
 \frac{\lambda_h}{4}\Big[ (1-\tth^2)^2 h_1^4 + 4\tth(1-\tth^2)h_1^3(h_2+v) 
	+ 4\tth^2 h_1^2 (h_2^2 + 2v h_2) \Big]   \nonumber \\[1pt]
&+& \frac{1}{4}\Big[(\cth^2\lambda_{h\phi}+\sth^2\lambda_{s\phi})h_1^2 
	+ (\sth^2\lambda_{h\phi}+\cth^2\lambda_{s\phi})(h_2^2 + 2v h_2)
	+ s_{2\theta}(\lambda_{h\phi} - \lambda_{s\phi})h_1(h_2+v) \Big]\phi_i^2 + \frac{1}{4}\lambda_\phi (\phi_i^2)^2
	,\phantom{XX}			
\eeqr
where $s_{2\theta} \equiv \sin 2\theta$.
Note that we have discarded some of the scalar interaction terms in the potential in Eq.~(\ref{eq:V_interaction}) 
by imposing the constraints in Eq.~(\ref{eq:lambdamin}).
For instance, 
the $h_1h_2^2$ interaction is absent because the relevant Higgs-scalar coupling 
$c_{122}\propto 1-t_\beta t_\theta$ vanishes under the constraint $t_\theta=1/t_\beta$.
Therefore, even if the radiatively generated $h_2$ mass is less than a half of the $h_1$ mass, 
the partial decay width $\Gamma\left(h_1 \to h_2 h_2\right)$ is negligible 
and our model is not constrained by the invisible Higgs decay measurements.

\section{Effective Potential}

The original approach of GW expressed the one-loop effective potential in terms of
the spherical-coordinate (radial) field of the scalar gauge eigenstates.
Rather differently, we derive the effective potential with the physical eigenstates of the scalars
and obtain the scalar masses at one-loop level directly.
Let the background value of the physical scalar $h_i$ be $h_{ic}$. 
Then the effective potential is obtained by expanding the interaction terms in the Lagrangian 
around the background fields $h_{ic}$
and by keeping terms quadratic in fluctuating fields only.
From Eq.~(\ref{eq:V_interaction}), the effective potential at one-loop level is given by
\beq
\Veff(h_{1c},h_{2c}) = V^{(0)}(h_{1c},h_{2c}) + V^{(1)}(h_{1c},h_{2c}),
\eeq
with
\beqr
V^{(0)}(h_{1c},h_{2c}) &=& \frac{\lambda_h}{4}\left[\left(1-\tth^2\right)^2h_{1c}^4
	+4\tth(1-\tth^2)h_{1c}^3h_{2c} + 4\tth^2h_{1c}^2h_{2c}^2 \right],  \nonumber \\[1pt]
V^{(1)}(h_{1c},h_{2c}) &=& 
	\sum_P n_P \frac{\bar{m}_P^4(h_{ic})}{64\pi^2}\left(\ln\frac{\bar{m}_P^2(h_{ic})}{\mu^2} - c_P\right),
\eeqr
where $c_P = 3/2\ (5/6)$ for scalars and fermions (gauge bosons) in the $\msbar$ scheme and
$\mu$ is a renormalization scale. 
$\bar{m}_P$ is a field-dependent mass and
the summation is over the particle species of fuctuating fields $P = h_{1,2}, Z, W^\pm, t, \phi_i$
and their degrees of freedoms ($n_P$) are given as follows
\beq
n_{h_1}=n_{h_2}=n_{\phi_i}=1,\quad n_Z=3,\quad n_{W^\pm}=6,
\quad n_t=-12.
\eeq
Taking the flat direction of the VEVs, 
we minimize the effective potential at $h_{1c} = 0$ and $h_{2c} = v$, 
which corresponds to $h_{c} = v_h$ and $s_{c} = v_s$ 
in terms of the background values of the scalar gauge eigenstates. 
The field-dependent mass $\bar{m}_P(h_{ic})$ is proportional to $h_{ic}$, 
so that $\bar{m}_P(h_{1c})$ is irrelevant to our study 
because $\partial \bar{m}_P(h_{1c})/\partial h_{1c} |_{h_{1c}=0}=0$.
The relevant field-dependent masses for $h_{2c}$ are obtained as 
\beqr
\bar{m}_{h_1}^2(h_{2c}) = 2\lambda_h\tth^2h_{2c}^2, &\quad&  \bar{m}_{h_2}^2(h_{2c}) = 0, \quad
\bar{m}_{\phi_i}^2(h_{2c}) = \half\left(\lambda_{h\phi}\sth^2 + \lambda_{s\phi}\cth^2\right)h_{2c}^2, \nonumber \\[1pt]
\bar{m}_{Z}^2(h_{2c}) =  M_Z^2\frac{h_{2c}^2}{v^2}, &\quad&
\bar{m}_{W^\pm}^2(h_{2c}) = M_W^2\frac{h_{2c}^2}{v^2},	\quad
\bar{m}_{t}^2(h_{2c}) = M_t^2\frac{h_{2c}^2}{v^2}.	
\eeqr
The GW scale $\Lambda$ can be obtained by applying the minimization condition of the effective potential,
and we have $\Lambda \simeq 0.85 M_\phi $ for $N=$2.
We exploit the numerical values of the physical observables at this scale.	

The masses of the physical scalars $h_{i}$ can be directly obtained by taking the second-order derivatives
of the effective potential with respect to the classical background fields $h_{ic}$ as
\beqr \label{eq:scalar_mass}
M_1^2 &=& \frac{\partial^2\Veff}{\partial h_{1c}^2}\Big{|}_{\substack{h_{1c}=0 \\ h_{2c}=v}} 
	= 2 \lambda_h v^2 \tth^2, 	\nonumber \\[1pt]
M_2^2 &=& \frac{\partial^2\Veff}{\partial h_{2c}^2}\Big{|}_{\substack{h_{1c}=0 \\ h_{2c}=v}} 
	= \frac{1}{8\pi^2v^2}\left(M_1^4 + 6M_W^4 + 3M_Z^4 -12M_t^4 + N M_{\phi}^4\right).
\eeqr
Although we have employed the strategy somewhat different from those of earlier studies following the GW approach
in Refs. \cite{Farzinnia:2014yqa, Ghorbani:2015xvz},
the final results for the scalar masses are equivalent. 
One can read off the inequality
$N M_{\phi}^4 \ge 12M_t^4 - M_1^4 - 6M_W^4 - 3M_Z^4$ from Eq.~(\ref{eq:scalar_mass})
and find that $M_{\phi} \gtrsim 265$ GeV for $N=2$.
In total, besides $N$, 
we have five independent model parameters relevant to DM phenomenology.  
The four model parameters $\lambda_h$, $v_s$, $\lambda_{h\phi}$, and $\lambda_{s\phi}$  
determine the masses $M_{1,2}$, $M_{\phi}$, and the mixing angle $\theta$, 
while the DM self coupling $\lambda_\phi$ is irrelevant to our study on the DM-SM interactions.
The dependency of the model parameters are 
\beq \label{eq:parameters}
v = \frac{v_h}{\sth}, \quad v_s = \frac{v_h}{\tth}, \quad \lambda_h = \frac{M_1^2\cth^2}{2v_h^2}, 
\quad \lambda_{hs} = -\frac{M_1^2\sth^2}{v_h^2}, \quad \lambda_{s} = \frac{M_1^2\sth^2\tth^2}{2v_h^2},
\quad \lambda_{s\phi} = \left(\frac{2M_{\phi}^2}{v_h^2} - \lambda_{h\phi}\right)\tth^2.
\eeq
Given the fixed Higgs mass $M_1$ and $v_h \simeq 246$ GeV, we constrain three independent new physics (NP) parameters 
by taking into account various theoretical considerations and experimental measurements in the next section.

\section{Results of Analysis}

In this section we turn to the phenomenological analysis of the model, especially with
the global $O(2)$ symmetry in the hidden sector. It corresponds to the case containing two exact 
copies of the DM. 
Using the conditions provided in Sec.\ III, 
we perform the numerical analysis by varying the following three NP
parameters: $\tth$, $M_{\phi}$, $\lambda_{h\phi}$.
Let us first consider the relic density. At present, the most accurate determination of the DM 
mass density $\Omega_{\rm DM}$ comes from global fits of cosmological parameters to a variety of 
observations such as measurements of the anisotropy of the cosmic microwave background (CMB) 
data by the Planck experiment and of the spatial distribution of galaxies \cite{Tanabashi:2018oca}:
\beq
\Omega_{\rm DM}h^2 = 0.1186 \pm 0.0020.
\label{eq:relic_obserb}
\eeq
This relic density observation will exclude some regions in the model parameter space.
The relic density analysis in this section includes all possible channels of
the $\phi_i \phi_i$ pair annihilation into the SM particles.  	
In this work, we implement the model described in Sec.\ II into the CalcHEP package \cite{Belyaev:2012qa}. 
By employing the numerical package \texttt{micrOMEGAs} \cite{Belanger:2018mqt} 
that includes the CalcHEP for computing the relevant annihilation cross sections, 
we compute the DM relic density and the spin-independent DM-nucleon scattering cross sections.

\begin{figure}[!hbt]
	\centering%
	\includegraphics[width=15cm]{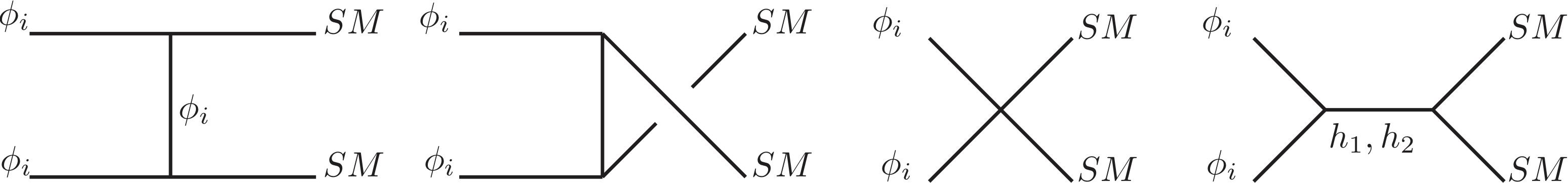}   
	\caption{Relevant Feynman diagrams for relic density calculation.}
	\label{diagram:relic}
\end{figure}

The relic density is determined by solving the Boltzmann equation, which contains the thermal 
average of the cross section times DM velocity $v_{\rm DM}$ as a proportional factor. By expanding 
this factor in powers of $v_{\rm DM}$, we find that 
\beq  \label{eq:DM_CS}
\langle \sigma v_{\rm DM} \rangle 
\propto
\frac{1}{4 M_{\phi}^2}\sqrt{1-\frac{M_{\rm SM}^2}{M_{\phi}^2}}
~\left\vert {\cal M} \right\vert^2 + {\cal O}\left( v_{\rm DM}^2 \right).
\eeq 
The relevant Feynman diagrams are shown in Fig.~\ref{diagram:relic}. When the final states 
are the SM particles other than Higgs-like bosons such as $h_1$ and $h_2$, the only contributions 
are from the $s$-channel diagrams exchanging $h_1$ and $h_2$. 
The amplitude can be expanded in powers of a small parameter $M_{\rm SM}^2/M_{\phi}^2$:
\beqr\label{17}
\left\vert {\cal M} \right\vert 
&\sim& v_h g_{\rm SM} 
\left\vert
\frac{\lambda_{h\phi}}{4 M_{\phi}^2}
+\frac{1}{16 M_{\phi}^4}
\big[ 
\lambda_{h\phi}\left( M_1^2 c_\theta^2+M_2^2 s_\theta^2 \right)
+\lambda_{s\phi}\left( M_2^2 -M_1^2 \right) c_\theta^2 
\big] + \cdots
\right\vert ,
\eeqr 
where $g_{\rm SM}$ indicates a coupling of $h_i$ and the SM particle interactions.
Note that the $\lambda_{s\phi}$ scales as $M_{\phi}^2$ as shown in Eq.~(\ref{eq:parameters}). 
Nevertheless, the expression in Eq.~(\ref{17}) demonstrates that the actual
contribution of $\lambda_{s\phi}$ to 
$\langle \sigma v_{\rm DM}\rangle$ first emerges only 
at the next-to-leading order in $M_{\rm SM}^2/M_\phi^2$. 
Thus, both of $\lambda_{h\phi}$ and $\lambda_{s\phi}$ scale similarly, and
as a result, those contributions to $\langle \sigma v_{\rm DM} \rangle$ scale like $\sim 1/M_\phi^6$.

Higgs-like boson pairs such as $h_i h_i$ can be created through various topologies 
including $s$-, $t$-, $u$-channels,
and contact interactions as shown in Fig.~\ref{diagram:relic}.
This is in contrast to the pure SM final states excluding Higgs-like particles that are
created only through the $s$-channel diagrams. 
For example, when the final states are a pair of $h_1$'s, 
the amplitude at leading order in $v_{\rm DM}$ and $M_{\rm SM}^2/M_{\phi}^2$ is approximately given as
\beq
\left\vert {\cal M} \right\vert \sim
\left\vert \left(s\, \textrm{channel}\right)- v_h^2 c_\theta^2 \frac{\left(\lambda_{h\phi}-\lambda_{s\phi}\right)^2}{M_{\phi}^2}
+\left(\lambda_{h\phi} c_\theta^2+\lambda_{s\phi} s_\theta^2 \right)+\cdots \right\vert .
\eeq
Substituting this $|\mathcal{M}|$ into Eq.~(\ref{eq:DM_CS}), we
obtain $\langle \sigma v_{\rm DM}\rangle$.
The leading contribution of $\lambda_{h\phi}$ to $\langle \sigma v_{\rm DM}\rangle$
scales as $1/M_\phi^2$ while that of $\lambda_{s\phi}$ scales as $M_\phi^2$.
As a result, as the DM mass increases,
the contributions of 
$\lambda_{h\phi}$ 
and $\lambda_{s\phi}$ to the relic density
vary in opposite directions:
the relic density is enhanced by
$\lambda_{h\phi}$-dependent interaction 
and 
is reduced by the $\lambda_{s\phi}$ counterpart.

\begin{figure}[!hbt]
	\centering%
	\includegraphics[width=8cm]{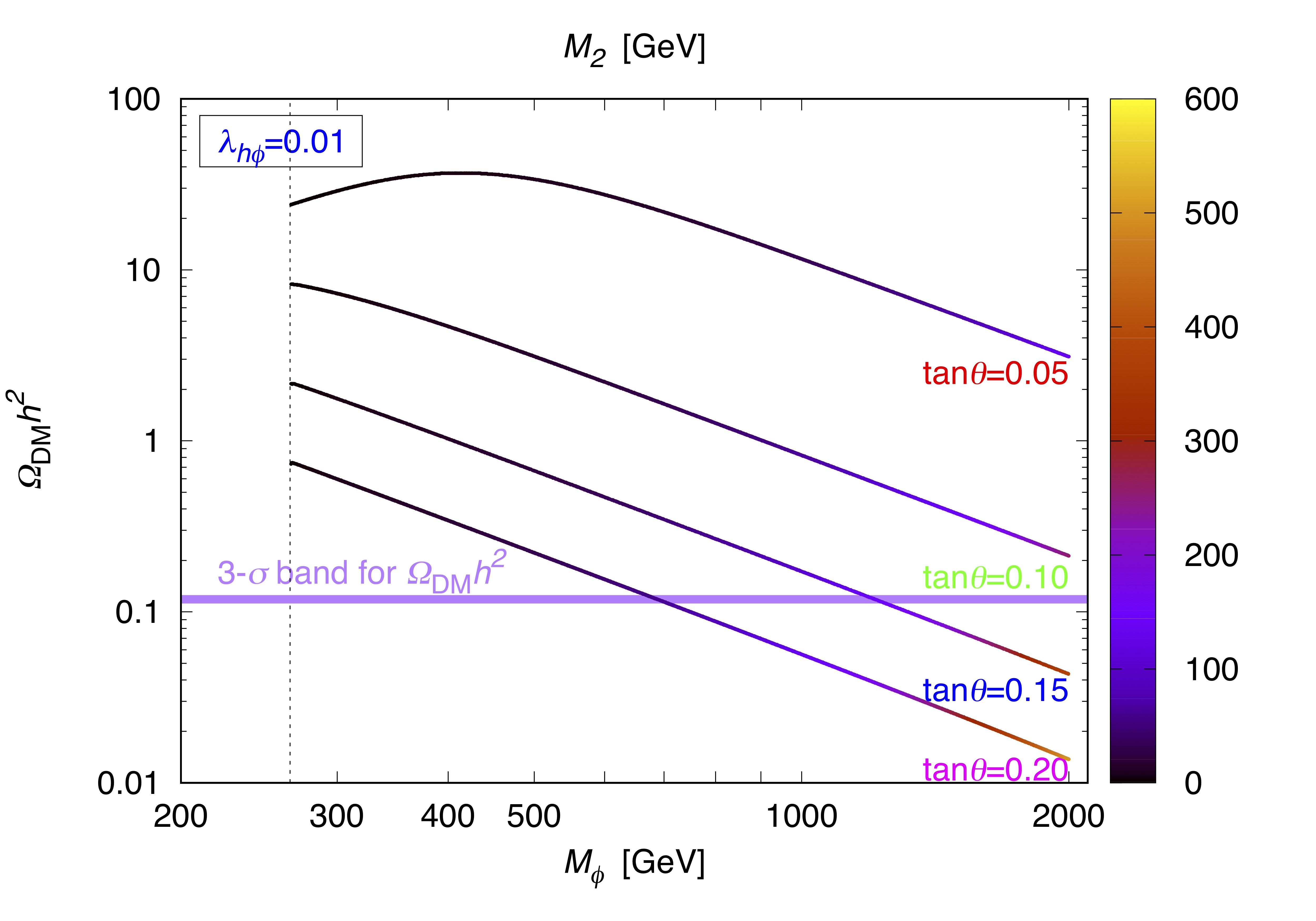}
	\includegraphics[width=8cm]{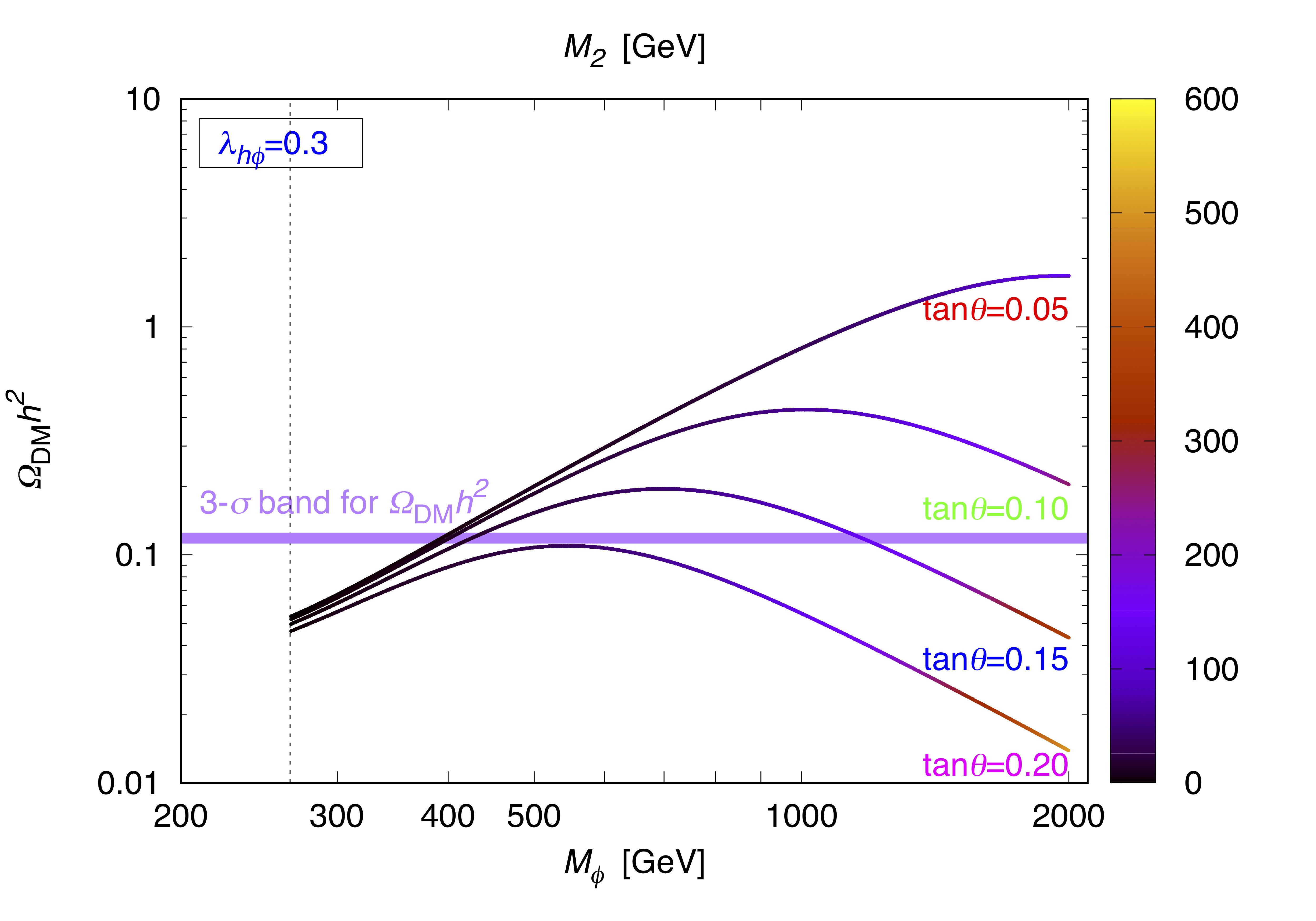}
	\caption{Relic density $\Omega_{\rm DM}$ as a function of the DM mass $M_{\phi}$, for 
		$\lambda_{h\phi}=0.01$\ (left panel) and 
		$\lambda_{h\phi}=0.3$\ (right panel). 
		The vertical dotted line indicates the minimum value of $M_\phi$ obtained in Eq.~(\ref{eq:scalar_mass})
		for $N=2$. }
	\label{DM-relic-lhp}
\end{figure}

The numerical results for the relic density are shown in Fig.~\ref{DM-relic-lhp}. 
We choose four different values of the mixing angle 
$\tth=$ 0.05, 0.1, 0.15, and 0.2 as benchmark points. 
The left panel of Fig.~\ref{DM-relic-lhp} shows the behavior of the relic density 
as a function of the DM mass for $\lambda_{h\phi}=0.01$. 
At this value of $\lambda_{h\phi}$,
the $\lambda_{h\phi}$ contribution is negligible unless $t_\theta$ is very small.
This generic suppression is originated from the scaling rule $\lambda_{s\phi}\sim t^2_\theta$ 
given in Eq.~(\ref{eq:parameters}).  
Thus, only for $t_\theta=$0.05 and for the small DM mass, $\lambda_{h\phi}$ contribution can compete 
with $\lambda_{s\phi}$'s and a local maximum appears. 
For other values of $t_\theta$, 
$\lambda_{s\phi}$ contribution dominates and relic density is a decreasing function of the DM mass 
$M_{\phi}$ as shown in the figure.
The right panel shows the case of  $\lambda_{h\phi}$=0.3, 
in which the $\lambda_{h\phi}$ contributions are not negligible at all. 
In that case, for small values of DM mass, $\lambda_{h\phi}$ contributions are dominant and 
the relic density is increasing as we increase the DM mass. In the case of large DM masses, 
$\lambda_{s\phi}$ is enhanced and its contributions make the relic density a decreasing 
function of the DM mass. Again, the mixing angle suppression of the $\lambda_{s\phi}$ makes 
the overturn of the relic density take place at larger DM mass when $t_\theta$ is smaller. 
The dotted vertical line represents the minimally allowed DM mass which is determined by 
the condition of non-tachyonic mass for $h_2$, the pseudo Goldstone boson of the spontaneously 
broken conformal symmetry of the model. If there is an $O(2)$ symmetry in the hidden sector, 
it means that there exist two exactly the same copies of the DM, 
and the minimum value of their masses is about $\sim 265$ GeV.

\begin{figure}[!hbt]
	\centering%
	\includegraphics[width=8.8cm]{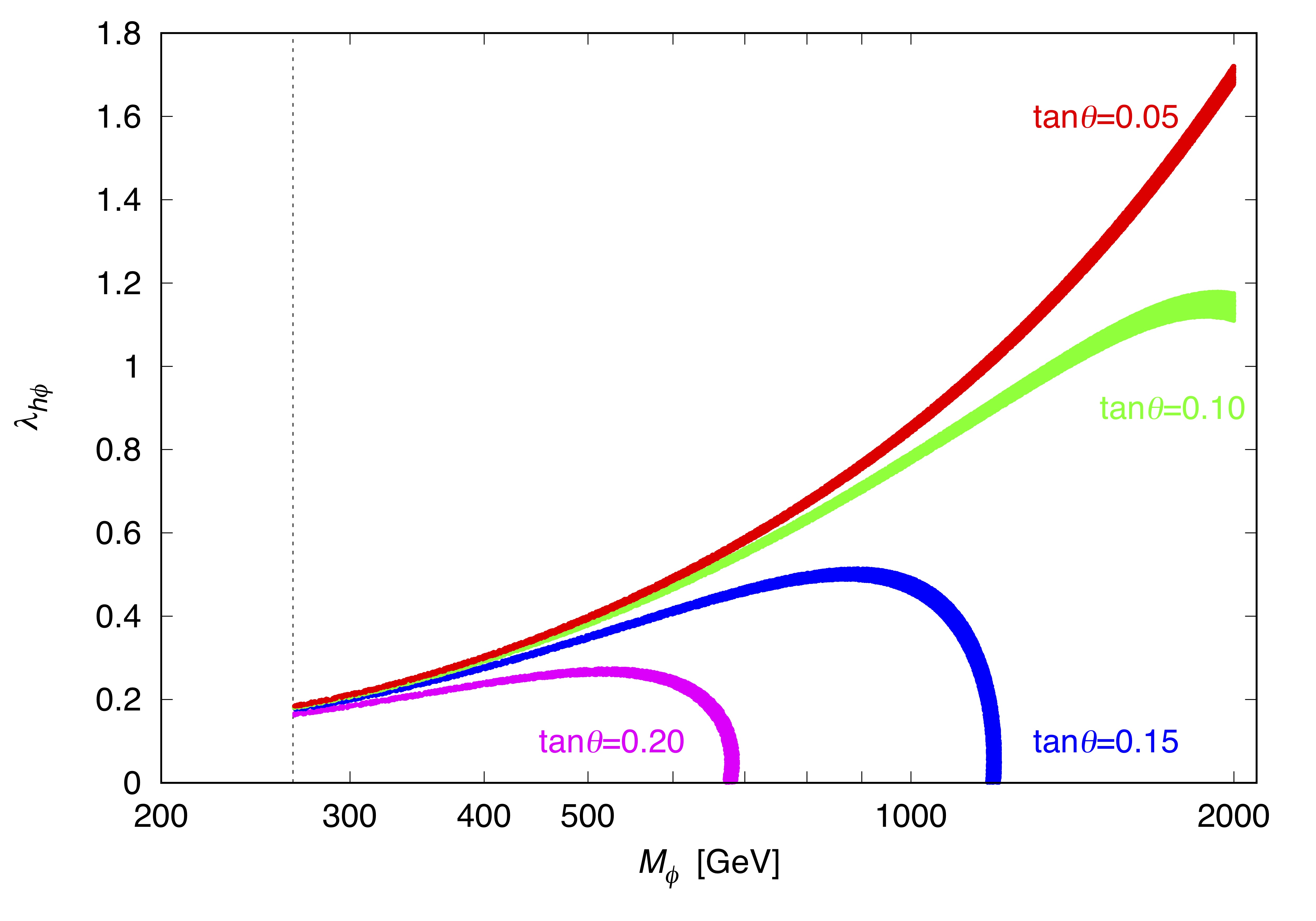}
	\caption{Allowed regions for the parameter set ($M_\phi$, $\lambda_{h\phi}$) by relic density observations 
		within $3\sigma$ range for four different values of $\tan\theta$. 
	The vertical dotted line indicates the minimum value of $M_\phi$ obtained in Eq.~(\ref{eq:scalar_mass})
	for $N=2$. }
	\label{DM-relic-sol}
\end{figure}

In Fig.~\ref{DM-relic-sol}, we show the parameter space satisfying the observed relic density 
in $3\sigma$ range for four different values of the mixing angles $t_\theta$=0.05, 0.1, 0.15 and 0.2.  
We have imposed the perturbativity constraints to
all of the dimensionless couplings in such a way that they should be smaller than $4\pi$, 
and the DM mass has been scanned up to $2$ TeV. Precise measurements 
of the relic density result in strong correlations between $\lambda_{h\phi}$ and the DM mass $M_{\phi}$, 
which is directly related to $\lambda_{s\phi}$, depending on the mixing angle $t_\theta$.  
Interestingly, there exists an upper bound for the DM mass for the given values of the mixing angle $\theta$.
Since $\lambda_{s\phi}$ has a factor $t_\theta^2$ as addressed in the previous paragraph, 
it is easy to see that the larger values of $\lambda_{h\phi}$ are disfavored by the 
observed relic density for large mixing angles. 
One can see from the figure that 
a relatively larger $\lambda_{h\phi}$ is only allowed for a smaller $\tth$
which results in the further suppression of $\lambda_{hs}$ and $\lambda_s$. 
As a result, the coupling $\lambda_h$ changes very slowly with the renormalization scale
as discussed earlier in the section II and so $\lambda_{h\phi}$ does.
Taken into account the direct detection bound which we will discuss in the next paragraphs,
the maximum possible value of $\lambda_{h\phi}(M_\phi)$ is about 1.6.
Neglecting other small contributions, 
we find that the dominant ones to the one-loop $\beta$-function of $\lambda_{h\phi}$ is given by
$4\pi^2\partial\lambda_{h\phi}/\partial\ln\mu \sim \lambda_{h\phi}(3\lambda_h + 4\lambda_{h\phi} + 2\lambda_{\phi})$ for $N=2$.
$\lambda_{\phi}$ is unknown and full analysis including $\lambda_{\phi}$ is beyond the scope of this study.
But for small $\lambda_{\phi} (\sim 0.01)$, 
our rough estimate is that Landau pole does not appear below the Planck scale for $\lambda_{h\phi}(M_\phi) \lesssim 0.5$.
For $0.5< \lambda_{h\phi}(M_\phi) < 1.6$, the coupling hits Landau pole above the scale larger than $10^9$ GeV,
but we do not discard this case and show its phenomenological implications 
on the following discussions as a reference.

\begin{figure}[!hbt]
	\centering%
	\includegraphics[width=10cm]{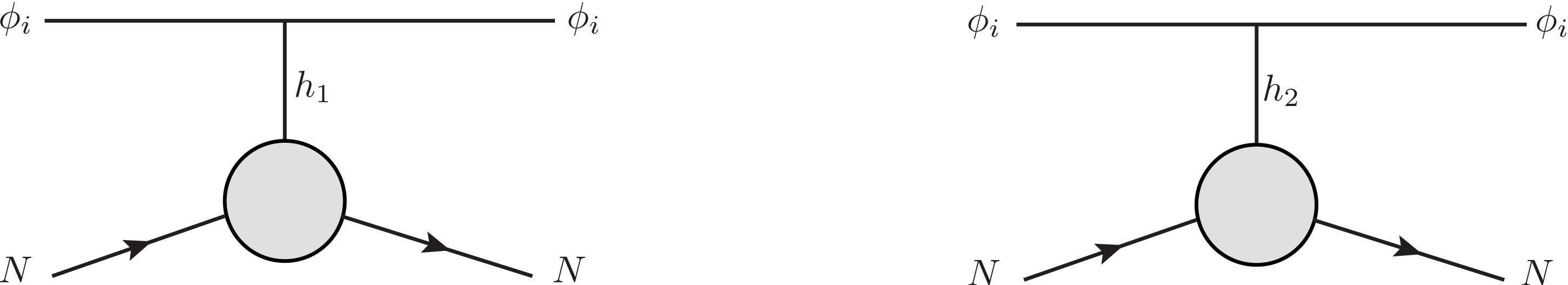}
	\caption{Relevant Feynman diagrams for spin-independent DM-nucleon scattering cross section.} 
	\label{diagram:SI}
\end{figure}

\begin{figure}[!hbt]
	\centering%
	\includegraphics[width=8cm]{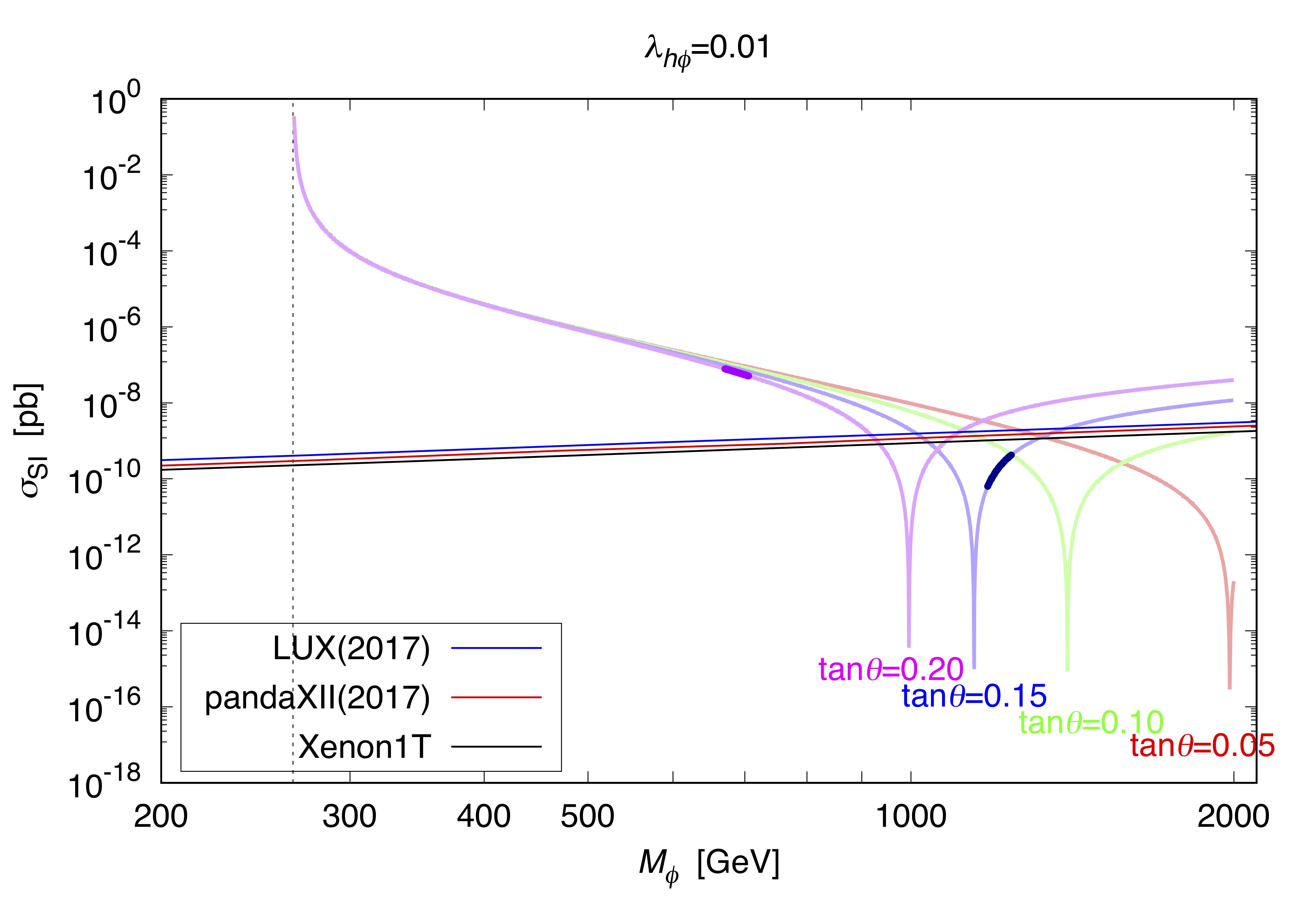}
	\includegraphics[width=8cm]{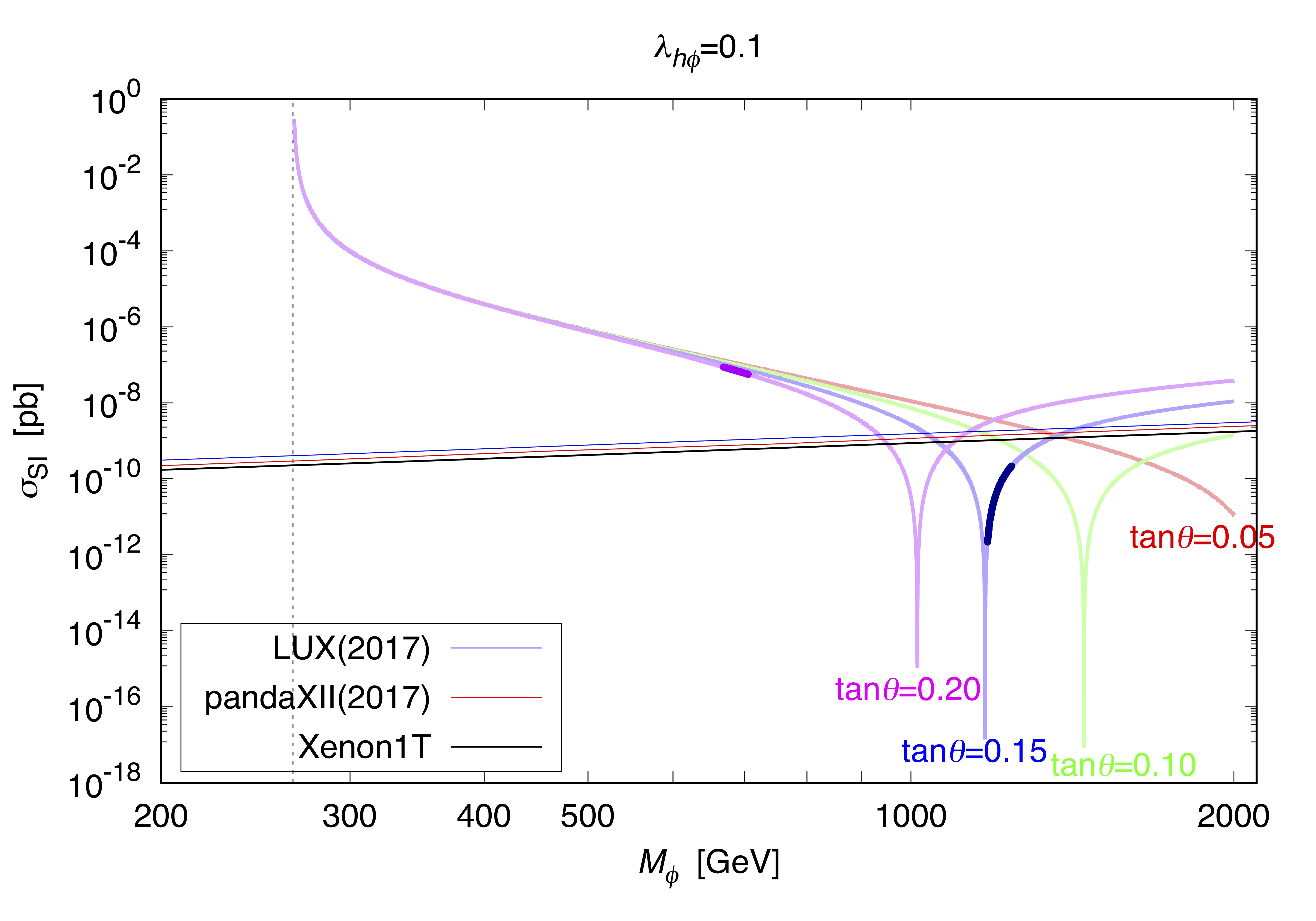}
	\includegraphics[width=8cm]{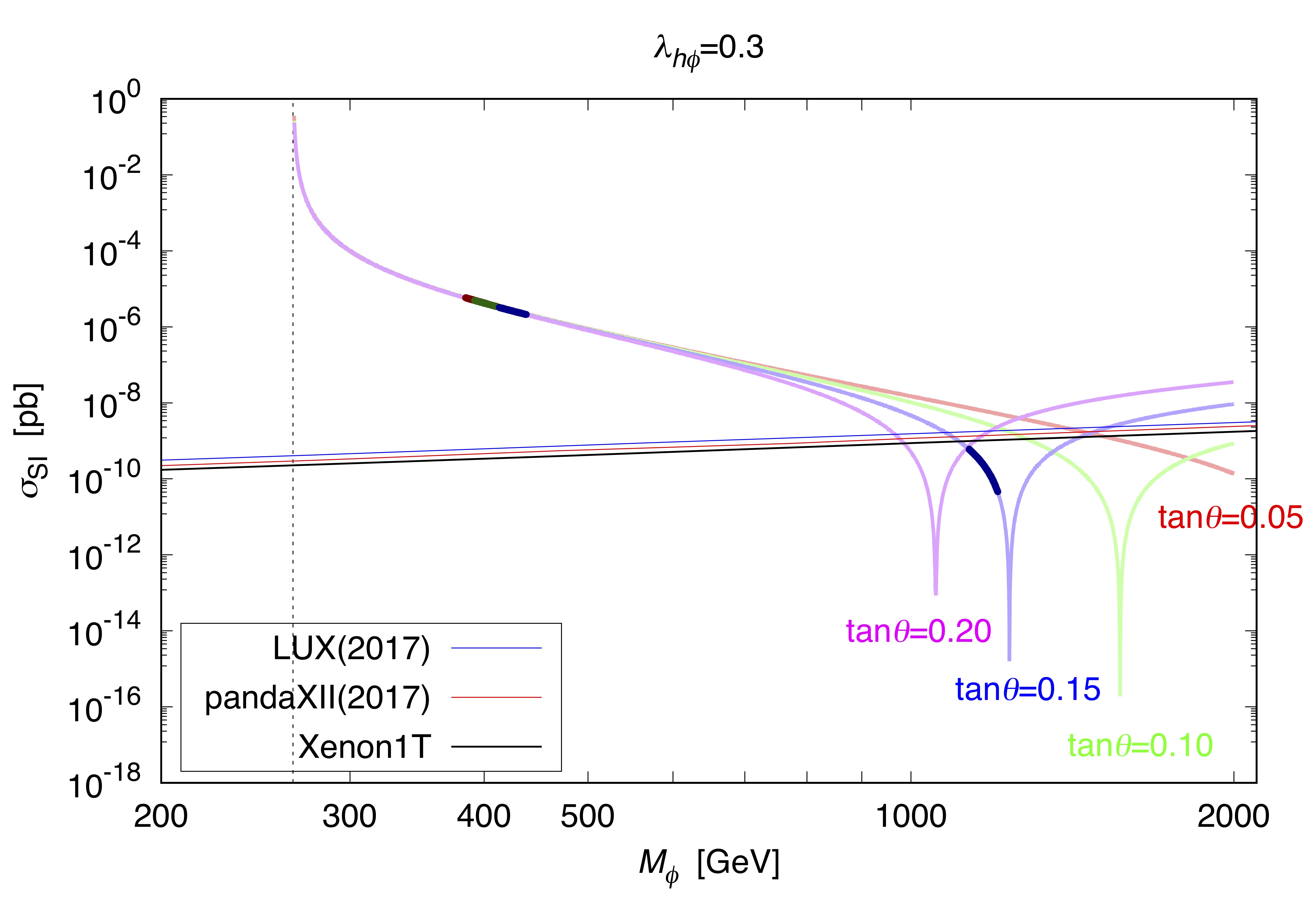}
	\includegraphics[width=8cm]{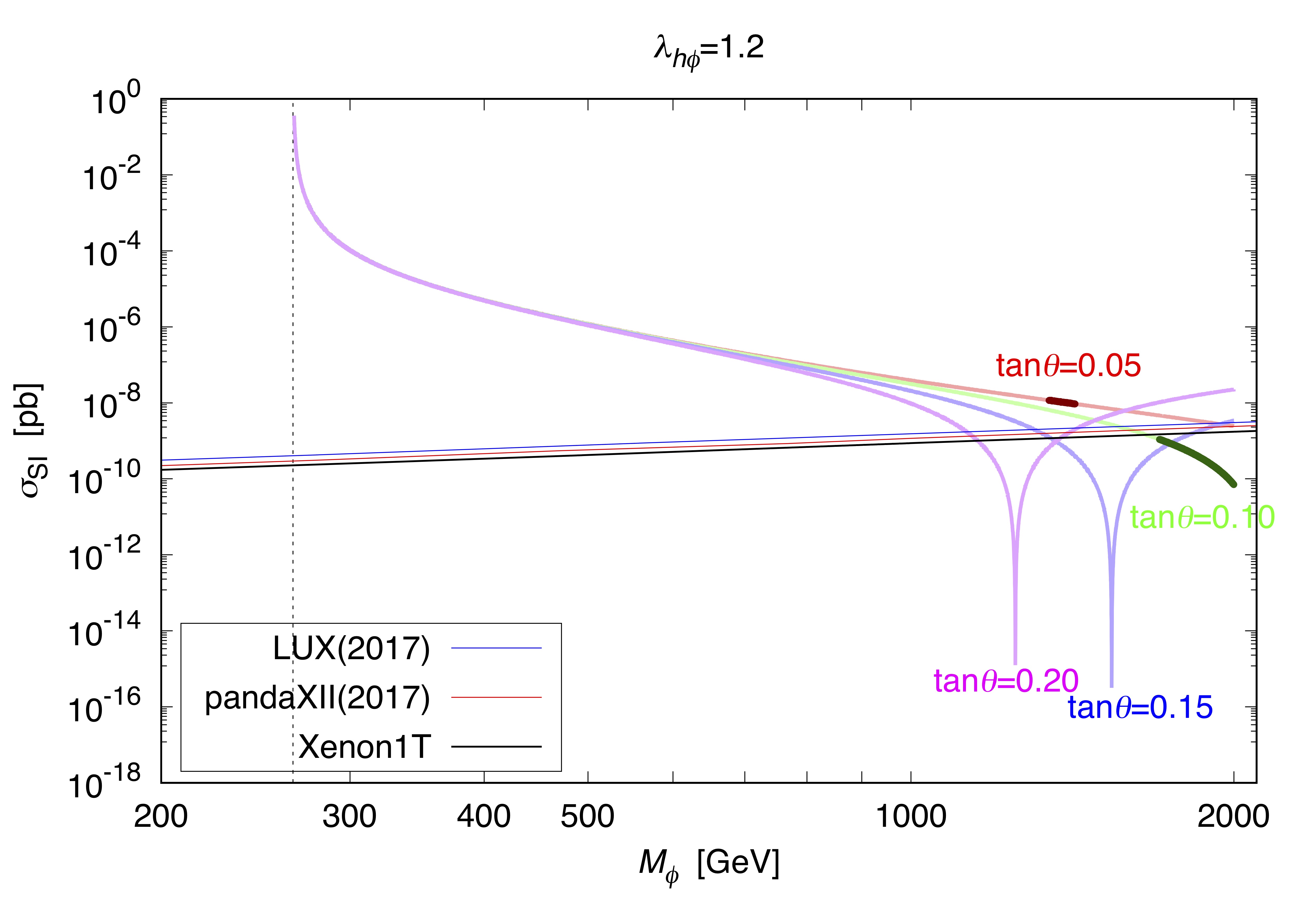}
	\caption{Spin-independent DM-nucleon scattering cross section as a function of DM mass.
		The thick bands are allowed by relic density observations, and	
		the vertical dotted line indicates the minimum value of $M_\phi$ obtained in Eq.~(\ref{eq:scalar_mass})
		for $N=2$. }
	\label{DM-SI-lhpxx}
\end{figure}

\begin{figure}[!hbt]
	\centering%
	\includegraphics[width=10cm]{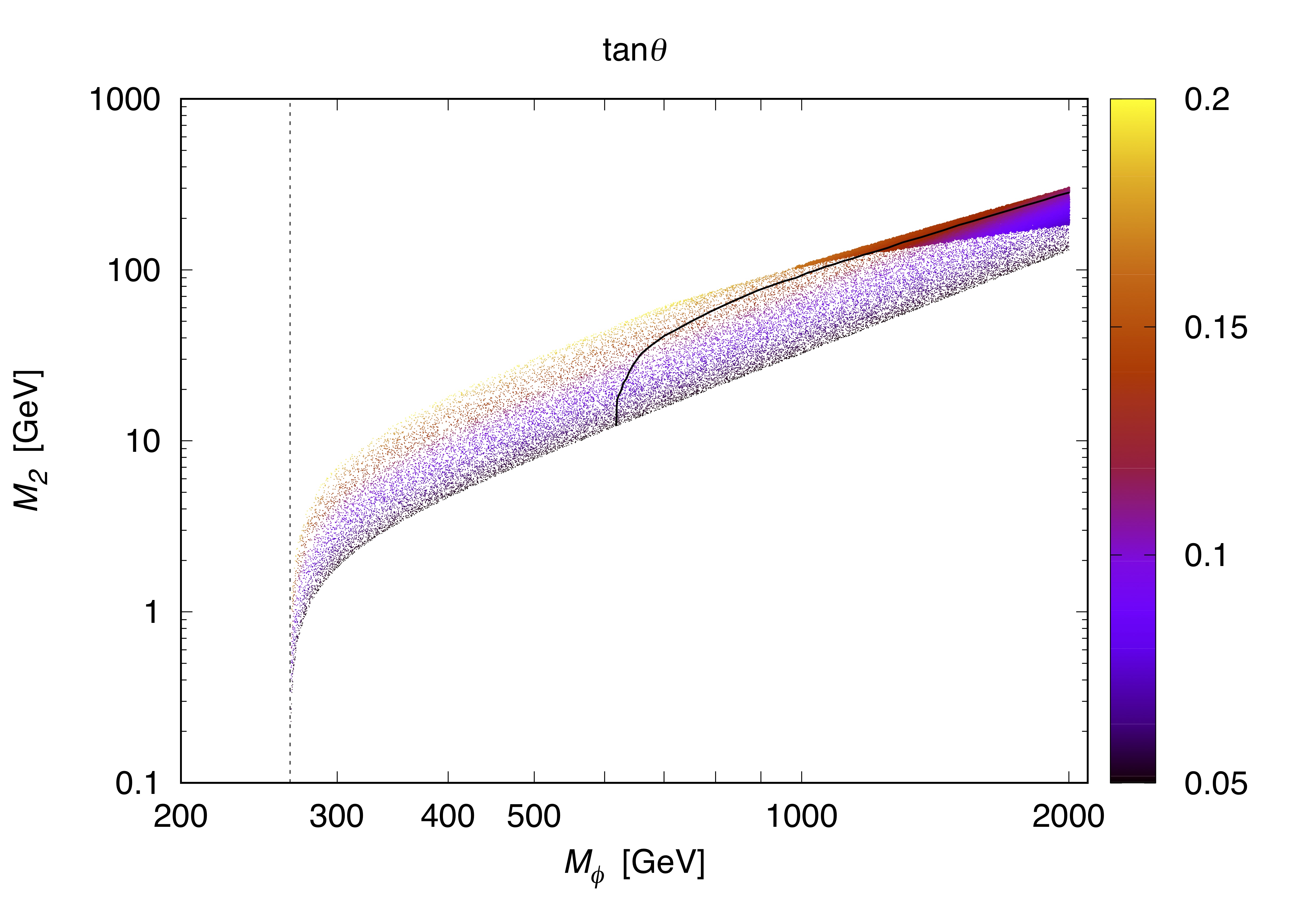}
	\caption{Allowed parameter sets of ($M_\phi$, $M_2$) by relic density observations in $3\sigma$ range. 
		Thick points are allowed by the direct detection bounds as well.
		The vertical dotted line indicates the minimum value of $M_\phi$ obtained in Eq.~(\ref{eq:scalar_mass}) for $N=2$. 
	   The allowed parameter sets above the solid line satisfy the bound $\lambda_{h\phi}(M_\phi) \leq 0.5$
       obtained from the Landau pole condition.  }
	\label{DM-h2}
\end{figure}

Next, let us consider the implications of the direct detection experiments on the model. 
Non-observation of DM-nucleon scattering events is interpreted as an upper bound on DM-nucleon cross section. 
The most stringent bound is given by Xenon1T experiment \cite{Aprile:2018dbl} in 2018. 
Around the same time, LUX \cite{Akerib:2016vxi} and PandaX-II \cite{Cui:2017nnn} 
collaborations reported similar but slightly less stringent results. 
The DM-nucleon scattering occurs only through the two $t$-channel diagrams exchanging
$h_1$ and $h_2$ that are shown in Fig.~\ref{diagram:SI}. 
The cross section for this elastic scattering of the highly non-relativistic
DM is well approximated by expanding the amplitude
in powers of the velocity $v_{\rm DM}$ of the DM:
\beqr \label{eq:directd}
\sigma \propto \frac{1}{M_{\phi}^2} \left\vert {\cal M} \right\vert^2, \quad
\left\vert {\cal M} \right\vert 
&\sim& 
g^{\rm SM}_{hNN} v_h 
\left\vert
\lambda_{h\phi} \left( \frac{s_\theta^2}{M_2^2}+\frac{c_\theta^2}{M_1^2} \right)
+\lambda_{s\phi} c_\theta^2 \left( \frac{1}{M_2^2}-\frac{1}{M_1^2} \right)
+\cdots \right\vert .
\eeqr
In the limit that $\lambda_{h\phi}$ is negligible and $M_2 \approx M_1$,
the cross section is severely suppressed because of a strong cancellation between the
two $t$-channel diagrams exchanging $h_1$ and $h_2$.
If $h_2$ becomes very light, then
the kinematic enhancement of the $h_2$ exchange diagram dominates over 
the contribution of the SM-like Higgs boson $h_1$.
In addition to that, $M_2^2$ is proportional to $\sth^2$, and $\lambda_{s\phi}$ to $\tth^2$, 
the mixing angle dependence disappears in the small DM mass limit. 
This property is clearly observed in Fig.~\ref{DM-SI-lhpxx}. 
As we increase the DM mass, $\lambda_{s\phi}$ increases as well
and $h_1$ contributions become more significant. 
The cross sections hit those minima when $M_2 \simeq M_1$ 
and increase again mainly due to $\lambda_{s\phi}$. 
Such a behavior is obviously revealed  in the first figure in Fig.~\ref{DM-SI-lhpxx} 
where $\lambda_{h\phi}=0.01$ is small enough to be neglected. 
When $\lambda_{h\phi}$ is large, a dip appears for slightly heavier $h_2$. 
This manifests itself in Fig.~\ref{DM-SI-lhpxx} in such a way that the locations of the dip is 
moved to the right, in the larger DM mass area for larger values of $\lambda_{h\phi}$.  
The points allowed by the relic density measurement is also presented as thick bands.
In Fig.~\ref{DM-h2}, we show the allowed parameter sets of $M_\phi$ and $M_2$ 
over the parameter space,
\beq
\lambda_{h\phi} \in \left[0,~4\pi\right],~~~~
M_{\phi}/{\rm GeV} \in \left[265,~2000\right],~~~~
t_\theta \in \left[0,~0.2\right]. 
\eeq
All of the points are consistent with the relic density measurements within $3\sigma$ range, 
and the thick region is allowed by the direct detection measurement bounds. 
Especially, the allowed parameter sets above the solid line satisfy the bound $\lambda_{h\phi}(M_\phi) \leq 0.5$
obtained from the Landau pole condition. 
This is our main prediction of the extra scalar mass $M_2$ and the mixing angle $\theta$,
depending on the DM mass $M_\phi$.
We also provide with a lower bound around $\sim 988$ GeV
for the DM mass $M_\phi$, which is valid for all $O(N)$ types of model with $N\geq2$.
Our numerical result does not give an upper bound on $M_\phi$,
but the allowed mixing angle $\tan\theta$ is highly constrained in the heavy $M_\phi$ region
if one applies the bound $\lambda_{h\phi}(M_\phi) \leq 0.5$. 

\begin{figure}[!hbt]
	\centering%
	\includegraphics[width=8cm]{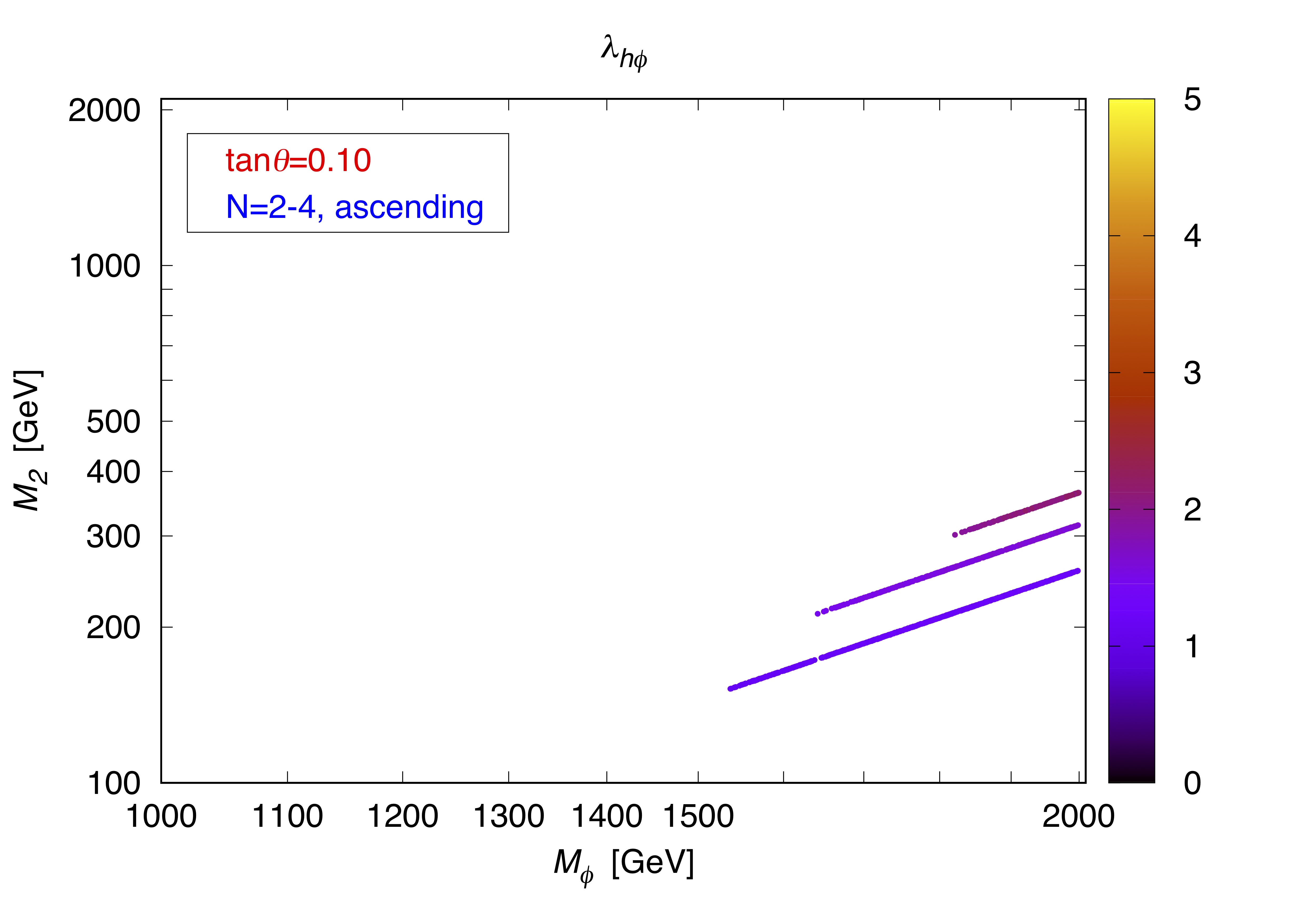}
	\includegraphics[width=8cm]{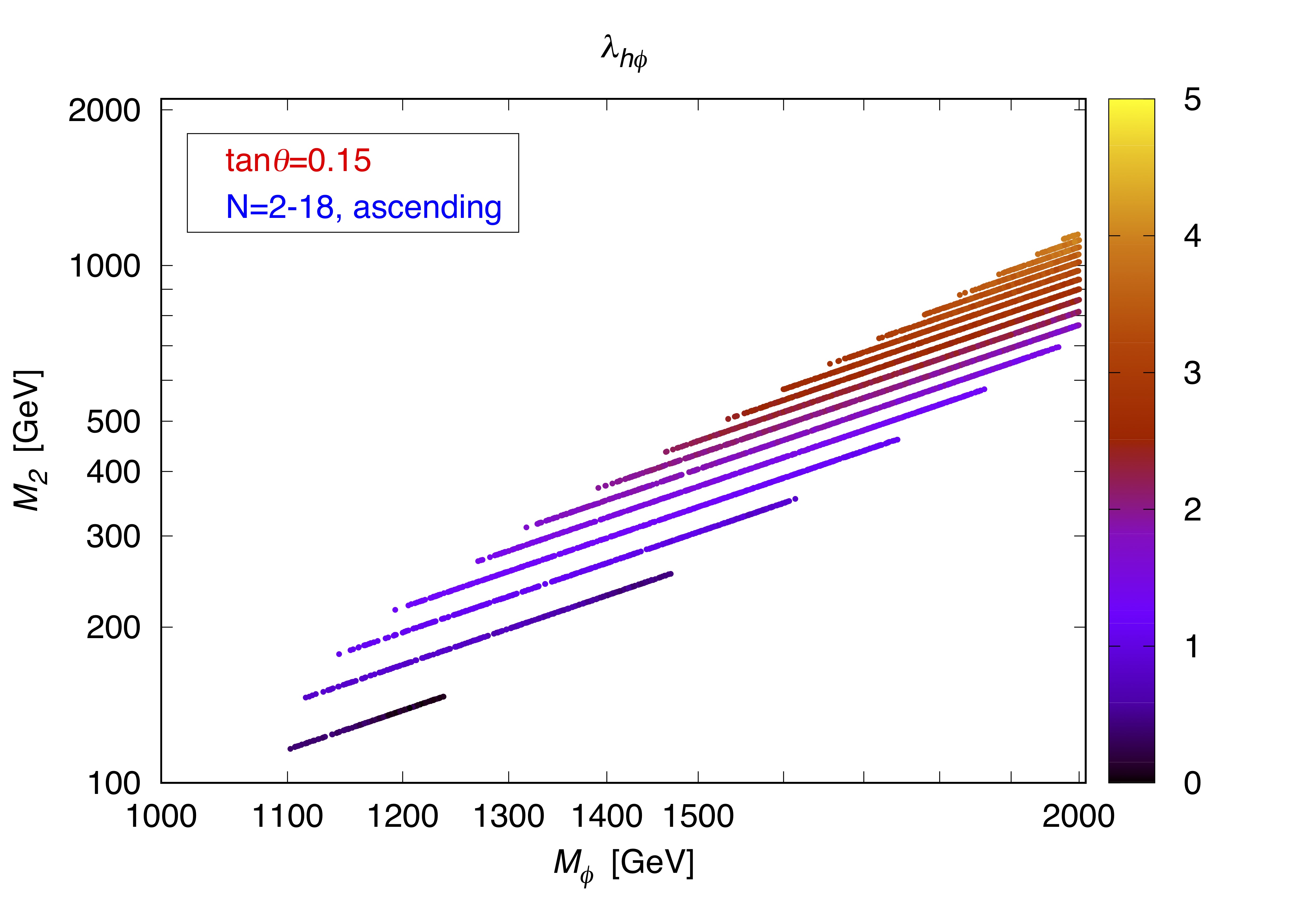}
	\caption{Solution points satisfying both relic density and direct detection constraints for 
		various values of $N$. There are no allowed parameter sets for $t_\theta \leq 0.05$. 
		The allowed region for $N \leq 3$ on the right panel satisfies the bound obtained from the Landau pole condition.} 
	\label{N-Sol}
\end{figure}

It is straightforward to extend our model to have $O(N)$ symmetry to stabilize the DM in the hidden 
sector. It corresponds to the case that there exist $N$ copies of the DM with the identical 
properties except that the minimum value for the DM mass $M_\phi$ can be lowered according to
Eq.~(\ref{eq:scalar_mass}). 
The results for the $O(N)$ extension are shown in Fig.~\ref{N-Sol} 
for $t_\theta=0.10$ (left) and $0.15$ (right).
One can see that a larger value of $\lambda_{h\phi}$ is required for a larger value of $N$ 
since the more we have the DM species, the smaller those portion to the relic density should be,
which means the larger annihilation cross section of one species of the DM. 
As a result, in the case of $t_\theta=0.15$ for instance,
phenomenologically allowed parameter sets do not satisfy the bound $\lambda_{h\phi}(M_\phi) \lesssim 0.5$
for $N \geq 4$.
In fact, for a larger $N$, the upper bound on $\lambda_{h\phi}(M_\phi)$ gets smaller.
Therefore, a small $N$ is preferred if the Landau pole condition is taken into account. 
This effect is demonstrated with the aid of color contours in  Fig.~\ref{N-Sol}. 
We also find that there are no allowed parameter sets for $t_\theta \leq 0.05$. 
If we increase the mixing angle, more allowed points are obtained but the mixing angle cannot be too large. 
For example, there are no allowed points for $N=2$ when $t_\theta=0.2$ 
with the DM mass up to $2$ TeV.

A few comments on other experimental constraints are in order. 
There are observational constraints on DM annihilation cross section 
such as Fermi-LAT \cite{Ackermann:2015zua} and  H.E.S.S. \cite{Abdallah:2016ygi} measurements.
They give stringent limits on the annihilation cross sections of the DM, especially on $b\bar{b}$ and $\tau\bar{\tau}$. 
But those constraints grow stronger for a lighter DM mass, around less than $800$ GeV,
while our model prefers DM mass heavier than about 1 TeV. 
Furthermore, in the high mass region, our model predicts far smaller cross sections 
than the experimental constraints.
More recently, Profumo $et\ al.$ \cite{Profumo:2017obk} suggested that the annihilation 
cross section of the DM to a mediator pair can be constrained by considering their successive
decays to SM particles. They showed that the upper bound of the annihilation cross sections 
must be $\sigma v_{\rm DM} \sim 10^{-25} \rm{cm}^3/s$ for the DM mass around 1 TeV. In our model, 
the DM annihilation to an $h_1$ pair was considered and its value lies around 
$\sim 10^{-26} \rm{cm}^3/s$ which is well below their bound when multiplied by the branching fractions
of the SM-like Higgs boson to SM particles, of which values is smaller than a few percent.
Since there is an extra scalar mediator in our model, we should also consider the observational 
constraint to the extra scalar particle. Big Bang Nucleosynthesis (BBN) gives a constraint 
on the lifetime of the extra scalar particle $h_2$ 
less than 1 second \cite{Jedamzik:2009uy,Kaplinghat:2013yxa},
and it is well satisfied in our model.  There are also constraints from the collider experiments. 
Our choice of the mixing angle $t_\theta \leq 0.2$ is quite safe against the LEP2 constraints 
since the $h_2$ mass is below around $300$ GeV for the DM mass up to $2$ TeV. 
Non-observation of Higgs-like particles in the high-mass Higgs searches through $WW$ and $ZZ$ modes
\cite{GonzalezLopez:2016xly,CMS:2016ilx,Angelidakis:2017ebh} at the LHC
also provides additional constraints under which our model still survives.
There are also constraints on the mixing angle from the signal strength measurements by 
ATLAS \cite{ATLAS:2019slw} and CMS \cite{CMS:2018lkl} collaborations, but 
they do not give severe restrictions to our analysis either.

Lastly, from the future collider experiments, 
one	might be able to probe the new scalar interaction effects directly or indirectly.
For instance, the new interactions modify the Higgs self-coupling $c_{111}$ for $h_1^3$ interaction sizably.
We found that the deviation of experimental value of $c_{111}$ from the SM expectation 
for $\tth \gtrsim  0.17$ lies within the expected precision of VLHC experiment, 
but not within HL-LHC precision \cite{Dawson:2013bba}. 
Also, the deviation of experimental value of Higgs boson couplings lies within the expected precision of ILC experiment
in $ZZ$ mode for $\tth \gtrsim 0.16$ at $\sqrt{s} = 250$ GeV 
and in both of $WW$ and $ZZ$ modes for $\tth \gtrsim 0.15$ at $\sqrt{s} = 500$ GeV, respectively \cite{Asner:2013psa}.
Therefore, $\tth \gtrsim 0.15$ case in this model can be tested in the future collider experiments.

\section{Summary and Conclusion}

In summary, we studied a classically scale-invariant DM model of scalar dark matters.
The model extends the Higgs sector to have an additional electroweak singlet scalar mediator, scalon,  
together with a scalar multiplet of global $O(N)$ symmetry, 
and the electroweak symmetry is broken via the CW mechanism.
The scalon serves as  the pseudo Nambu–Goldstone boson of scale symmetry breaking, and 
the scalar multiplet $\phi$ can be the viable DM candidate.
The DM scalar $\phi$ couples directly to the SM Higgs with the coupling $\lambda_{h\phi}$
which plays an important role in DM phenomenology.  

With the most general conformally invariant Lagrangian, 
we employed the framework of GW, where a flat direction at tree level is lifted by radiative corrections. 
Through the mixing of the scalon with the SM-like Higgs boson, 
two light scalar particles $h_1$ and $h_2$ emerge in the model
that interact with both visible and hidden sectors. 
After EWSB, the SM Higgs $h_1$ and the DM scalars $\phi$ have the tree-level masses 
while the new scalar singlet $h_2$ acquires its mass through radiative corrections
of the SM particles and $\phi$ as obtained in Eq.~(\ref{eq:scalar_mass}), 
so that $h_2$ mass $M_2$ is dependent on the scalar mixing angle $\theta$ and the DM mass $M_\phi$.

With three independent new parameters $\tth$, $M_\phi$, and $\lambda_{h\phi}$, 
we presented the allowed region of NP parameters satisfying the recent measurement of the relic abundance 
in Fig.~\ref{DM-relic-sol} for $N=2$ case.
We also showed the spin-independent DM-nucleon scattering cross section of the scalar DM 
by varying the DM mass $M_\phi$ with parameter sets allowed by the relic density observation, 
and compare the results with the observed upper limits from various experiments in Fig.~\ref{DM-SI-lhpxx}.
In the figures, one can clearly see that
the allowed parameter space constrained by the relic density observation are located 
in the mass region of $M_\phi$ heavier than about 1 TeV.
We performed the numerical analysis for $\tth \leq 0.2$, and the $\tth \leq 0.05$ case is disfavored 
in this model due to the recent Xenon1T bound.
For $N > 2$, we showed in Fig.~\ref{N-Sol}
that having a too large value of $N$ is disfavored especially due to the direct detection bounds.
We also found that our model is not constrained by the current indirect detection bounds for the given parameter sets,
and $\tth \gtrsim 0.15$ case can be tested in the future collider experiments. 
This model can be expanded by introducing complex scalars and/or a gauge symmetry in the hidden sector as well
if necessary.
Furthermore, our model can possibly resolve the unexplained anomalies in the CMB, 
so-called the small-scale problems in galaxy formation, due to the existence of the DM self coupling $\lambda_{\phi}$,
and it will be given in our forthcoming studies.

\acknowledgments
This work was supported by Basic Science Research Program
through the National Research Foundation of Korea (NRF)
funded by the Ministry of Science, ICT and Future Planning under the 
Grant No. NRF-2017R1E1A1A01074699, 
and funded by the Ministry of Education under the Grant Nos. NRF-2016R1A6A3A11932830 (S.-h. Nam) 
and NRF-2018R1D1A1B07047812 (D.-W. Jung).

\def\npb#1#2#3 {Nucl. Phys. B {\bf#1}, #2 (#3)}
\def\plb#1#2#3 {Phys. Lett. B {\bf#1}, #2 (#3)}
\def\prd#1#2#3 {Phys. Rev. D {\bf#1}, #2 (#3)}
\def\jhep#1#2#3 {J. High Energy Phys. {\bf#1}, #2 (#3)}
\def\jpg#1#2#3 {J. Phys. G {\bf#1}, #2 (#3)}
\def\epj#1#2#3 {Eur. Phys. J. C {\bf#1}, #2 (#3)}
\def\arnps#1#2#3 {Ann. Rev. Nucl. Part. Sci. {\bf#1}, #2 (#3)}
\def\ibid#1#2#3 {{\it ibid.} {\bf#1}, #2 (#3)}
\def\none#1#2#3 {{\bf#1}, #2 (#3)}
\def\mpla#1#2#3 {Mod. Phys. Lett. A {\bf#1}, #2 (#3)}
\def\pr#1#2#3 {Phys. Rep. {\bf#1}, #2 (#3)}
\def\prl#1#2#3 {Phys. Rev. Lett. {\bf#1}, #2 (#3)}
\def\ptp#1#2#3 {Prog. Theor. Phys. {\bf#1}, #2 (#3)}
\def\rmp#1#2#3 {Rev. Mod. Phys. {\bf#1}, #2 (#3)}
\def\zpc#1#2#3 {Z. Phys. C {\bf#1}, #2 (#3)}
\def\cpc#1#2#3 {Chin. Phys. C {\bf#1}, #2 (#3)}

\end{document}